\documentstyle[12pt,epsfig]{article}

\def\be{\begin{equation}}
\def\ee{\end{equation}}
\def\ba{\begin{eqnarray}}
\def\ea{\end{eqnarray}}
\def\ga{\mathrel{\raise.3ex\hbox{$>$\kern-.75em\lower1ex\hbox{$\sim$}}}}
\def\la{\mathrel{\raise.3ex\hbox{$<$\kern-.75em\lower1ex\hbox{$\sim$}}}}

\newcommand{\fr}[2]{\frac{#1}{#2}}

\newcommand{\tot}{\rm{tot}}
\newcommand{\rel}{\rm{rel}}
\newcommand{\PRD}{Phys.\ Rev.\ D}
\newcommand{\AST}{astro-ph}
\begin{document}

\baselineskip=16pt
\begin{titlepage}
\rightline{astro-ph/0601333} \rightline{January 2006}
\begin{center}

\vspace{0.5cm}

\large {\bf Constraints on the coupled quintessence from cosmic
microwave background anisotropy and matter power spectrum}
\vspace*{5mm} \normalsize

{\bf Seokcheon Lee$^{\,1}$},  {\bf Guo-Chin Liu,$^{\,2}$ and
{\bf Kin-Wang Ng$^{\,1,2}$}}

\smallskip
\medskip

$^1${\it Institute of Physics,\\ Academia Sinica,
 Taipei, Taiwan 11529, R.O.C.}

$^2${\it Institute of Astronomy and Astrophysics,\\
 Academia Sinica, Taipei, Taiwan 11529, R.O.C.}

\smallskip
\end{center}

\vskip0.6in

\centerline{\large\bf Abstract} We discuss the evolution of linear
perturbations in a quintessence model in which the scalar field is
non-minimally coupled to cold dark matter. We consider the effects
of this coupling on both cosmic microwave background temperature
anisotropies and matter perturbations. Due to the modification of
the scale of cold dark matter as $\rho_{c} = \rho_{c}^{(0)} a^{-3
+ \xi}$, we can shift the turnover in the matter power spectrum
even without changing the present energy densities of matter and
radiation. This can be used to constrain the strength of the
coupling. We find that the phenomenology of this model is
consistent with current observations up to the coupling power
$n_{c} \leq 0.01$ while adopting the current parameters measured
by WMAP. Upcoming cosmic microwave background observations
continuing to focus on resolving the higher peaks may put strong
constraints on the strength of the coupling.

\vspace*{2mm}

\end{titlepage}

\section{Introduction}
\setcounter{equation}{0}

 Analysis of the Hubble diagram of high redshift Type Ia supernovae
(SNe Ia) has discovered that the expansion of the Universe is
currently accelerating \cite{SCP}. In addition, combining
measurements of the acoustic peaks in the angular power spectrum
of the cosmic microwave background (CMB) anisotropy which indicate
the flatness of the Universe \cite{CMB} and the matter power
spectrum of large scale structure (LSS) which is inferred from
galaxy redshift surveys like the Sloan Digital Sky Survey (SDSS)
\cite{SDSS} and the $2$-degree Field Galaxy Redshift Survey
($2$dFGRS) \cite{2dFGRS} has confirmed that a component with
negative pressure (dark energy) should be added to the matter
component to make up the critical density today.

 The cosmological constant and/or a quintessence field are the most
commonly accepted candidates for dark energy. The latter is a
dynamical scalar field leading to a time dependent equation of
state, $\omega_{\phi}$. Also, this scalar field has a fluctuating,
inhomogeneous component in order to conserve the equivalence
principle corresponding to the response of the new component to
the inhomogeneities in the surrounding cosmological fluid
\cite{CDS}.

 Several new observational effects produced by the existence of
quintessence are imprinted in the CMB anisotropies and the matter
power spectrum when compared to models with the cosmological
constant. The locations of the acoustic peaks in the CMB angular
power spectrum are shifted due to their dependence on the amount
of dark energy today and at last scattering as well as
$\omega_{\phi}$ \cite{Doran, SLee}. Usually the Universe is
dominated by the quintessence at late times ($z \sim {\cal
O}(1)$), when the gravitational potential associated with the
density perturbations is changed due to a time dependent
$\omega_{\phi}$. This enhances the CMB anisotropies at large
angular scales by the integrated Sachs-Wolfe effect (ISW)
\cite{ISW}. Thus the amplitudes of both CMB angular and matter
power spectra decrease at large scale for the fixed Cosmic
Background Explorer Satellite (COBE) normalization compared to those
in the cold dark matter model with a cosmological constant
($\Lambda$CDM)~\cite{Dodelson1}.

 The possibility that a scalar field at early cosmological times
follows an attractor-type solution and tracks the evolution of the
visible matter-energy density has been explored \cite{Ratra}. This
may help alleviate the severe fine-tuning associated with the
cosmological constant problem. However, this still cannot explain
the reason why the dark energy and the dark matter have comparable
energy densities at present. Recently models considering the
coupling of quintessence to dark matter have been investigated as
a possible solution for this late time coincidence problem
\cite{coupQ}. However, several of these models using a simple
coupling can be ruled out by observational constraints
\cite{coupQ1}.

 These non-minimally coupled quintessence models have several different
observational effects compared to the minimally coupled models.
One of the most important effects is a different scaling of the
cold dark matter (CDM) compared to that of CDM of the minimally
coupled case. Since CDM scales as $\rho_{c} = \rho_{c}^{(0)} a^{-3
+ \xi}$ where $\xi < 0$, there will be more CDM energy density at
early epoch compared to the case with $\xi = 0$ ({\it i.e.} $n _c
=0$). As the coupling is increased, the locations of the acoustic
peaks are also shifted to smaller scales. However, the amplitudes
of the odd-number peaks decrease due to the decrease of the baryon
density at early time and the increasing ISW effect as the
coupling scaled with the COBE normalization. The amplitudes of the
even-number peaks increase due to the increase of the CDM energy
density. Even though the ratios of the height of the first peak to
those of higher peaks between different couplings are quite
similar to one another, these are quite different from the
$\Lambda$CDM model \cite{KMT}. The location of the turnover in the
matter power spectrum corresponds to the scale that entered the
Hubble radius when the universe became matter-dominated
($a_{eq}$). Thus the coupling might shift the turnover scale.

 This paper is organized as follows. In the next section we show
the basic equations of linear perturbations of the coupled
quintessence model. In Sec. 3, we derive the formal entropy
perturbation due to multiple fluids. We also consider the
possibility of isocurvature perturbations of the quintessence. We
check the coupling effects on CMB and matter power spectrum in
Sec. 4. The effect of coupling on the metric perturbation is
considered in Sec. 5. Our conclusion is in the last section.

\section{Linear perturbations}
\setcounter{equation}{0}

We will consider the perturbation effect of the quintessence model
which is coupled to the CDM. First we start from the
metric in the conformal Newtonian (longitudinal) gauge, which is
restricted only for the scalar mode of the metric perturbations
\cite{Perturb}. The line element is given by \be ds^2 = a^2(\eta)
\Biggl[ -\Bigl(1 + 2 \Psi(\eta, \vec{x}) \Bigr) d\eta^2 + \Bigl(1
- 2 \Phi(\eta,\vec{x}) \Bigr) dx^i dx_i \Biggr], \label{CNG} \ee
where $\eta$ is the conformal time, $\Psi$ is the amplitude of
perturbation in the lapse function, and $\Phi$ is the amplitude of
perturbation of a unit spatial volume. We will consider only a
flat universe case. If the coupling is derived by a Brans-Dicke
Lagrangian, the radiation is decoupled from the dark energy
\cite{Amendola0}. Due to the strong constraint on the coupling to
the baryons from the local gravity experiments such as radar
time-delay measurements, we will assume that the baryons are
decoupled from quintessence \cite{Damour1}. Also from the
violation of the weak equivalence principle, even though it is
in a still
 way that is locally unobservable, we can have the
species-dependent couplings \cite{Damour2}. So we will make an
assumption that the scalar field $\phi$ is coupled only to CDM by
means of a general function $\exp[B_{c}(\phi)]$ and there is no
coupling to
 the baryons or the radiation \cite{Coupled}.
We can write the
 general equation including this interaction as
\be S = - \int d^4x
\sqrt{-g} \Biggl\{ \frac{\bar{M}^2}{2} [\partial^{\mu} \phi
\partial_{\mu} \phi - R] + V(\phi) - {\cal L}_{c}  - {\cal L}_{r}
- {\cal L}_{b} \Biggr\}, \label{lagrangian} \ee where $\bar{M} =
1/ \sqrt{8\pi G}$ is the reduced Planck mass and ${\cal L}_{i}$s
denote respectively CDM, radiation, and baryons. If we regard the
matter as a gas of pointlike particles with masses $m_{c}$ and
paths $x_{c}^{\nu}(t)$, then we can write ${\cal L}_{c}$ as \be
 {\cal L}_{c} = - \frac{m_{c}}{\sqrt{-g}} \delta(\vec{x} -
\vec{x}_{c}(t)) (- g_{\mu\nu} \dot{x}_{c}^{\mu}
\dot{x}_{c}^{\nu})^{1/2}, \label{Lm} \ee where $m_{c} =
e^{B_{c}(\phi)} m^{*}_{c}$ ($m_{c}^{*}$ being a bare mass of the
CDM) ~\cite{Peebles,LOP}. Each fluid element has an
energy-momentum tensor $T^{\mu}_{(\beta) \nu}$ where $\beta$
includes all of the species. The total energy-momentum tensor is
covariantly conserved, however
 the
energy-momentum transfer between CDM and quintessence is written
 as
\ba \sum_{\beta} \nabla_{\mu} T^{\mu}_{(\beta) \nu} &=& 0, \label{dTbeta} \\
\nabla_{\mu} T^{\mu}_{(\gamma) \nu} &=& 0, \label{dTgamma} \\
\nabla_{\mu} T^{\mu}_{(d) \nu} &=& Q_{(d) \nu}, \label{dTd} \ea
where $\gamma$ denotes baryons or radiation,
$d$ denotes CDM or quintessence, and $Q_{(d) \nu}$ is the
energy-momentum transfer vector which shows the energy transfer to
the $d$-fluid \cite{Kodama}. This transfer vector is constrained
as \be \sum_{d} Q_{(d) \nu} = 0. \label{sumQ} \ee From the above
action (\ref{lagrangian}) we have the following equation which
will give the constraint equation for the interaction between the
quintessence and the CDM: \be \bar{M}^2 \Box \phi - \frac{\partial
V}{\partial \phi} - \frac{\partial B_{c}}{\partial \phi} {\cal
L}_{c} = 0. \label{boxphi} \ee From this we can find the unperturbed part of
the field equation, \be \phi'' + 2 {\cal H}
\phi' + \frac{a^2}{\bar{M}^2} \frac{\partial V(\phi)}{\partial
\phi} = - \frac{a^2}{\bar{M}^2} \frac{\partial
B_{c}(\phi)}{\partial \phi} \rho_{c}, \label{phieq} \ee where the prime
means $d/d\eta$, ${\cal H} = a'/a$, and $\rho_{c}$ means the energy
density of CDM. As we mentioned before the energy-momentum of each
species may not be conserved due to the scalar field coupling even
though the total energy momentum does conserve. If we use this
fact, then we can rewrite Eqs.~(\ref{dTbeta}),
(\ref{dTgamma}), and (\ref{dTd}) as
\ba \rho_{\tot}' &=& - 3 {\cal H} (\rho_{\tot} + p_{\tot}), \label{rhotot'} \\
\rho_{\gamma}' &=& - 3 {\cal H} (\rho_{\gamma} + p_{\gamma}), \label{rhogamma'} \\
\rho_{c}' &=& - 3 {\cal H} (\rho_{c} + p_{c}) + B_{c,\phi} \phi'
\rho_{c} \equiv -3 {\cal H}(\rho_{c} + p_{c})(1 - {\cal B}_{c}), \label{rhoc'} \\
\rho_{\phi}' &=& -3 {\cal H}(\rho_{\phi} + p_{\phi}) - B_{c,\phi}
\phi' \rho_{c} \equiv -3 {\cal H}(\rho_{\phi} + p_{\phi})(1 -
{\cal B}_{\phi}), \label{rhophi'} \ea where $B_{c,\phi} = \partial
B_{c}/ \partial \phi$. Henceafter we will adopt the potential and the coupling
as
 given in Ref.~\cite{LOP}: \ba V(\phi) &=& V_{0} \exp \Bigl(
\frac{\lambda\phi ^2 }{2} \Bigr), \label{V} \\ \exp [B_{c}(\phi)]
&=& \Biggl(\frac{b_c+V(\phi)/V_0}{1+b_c}\Biggr)^{n_c}, ~~~ {\rm
with}~~ b_c+1>0, \label{BF} \ea
where $V_0$, $\lambda$, $b_c$, and $n_c$ are constant parameters.
Here we will consider only $b_{c} =0$ case.
\begin{center}
\begin{figure}
\vspace{1cm} \centerline{\psfig{file=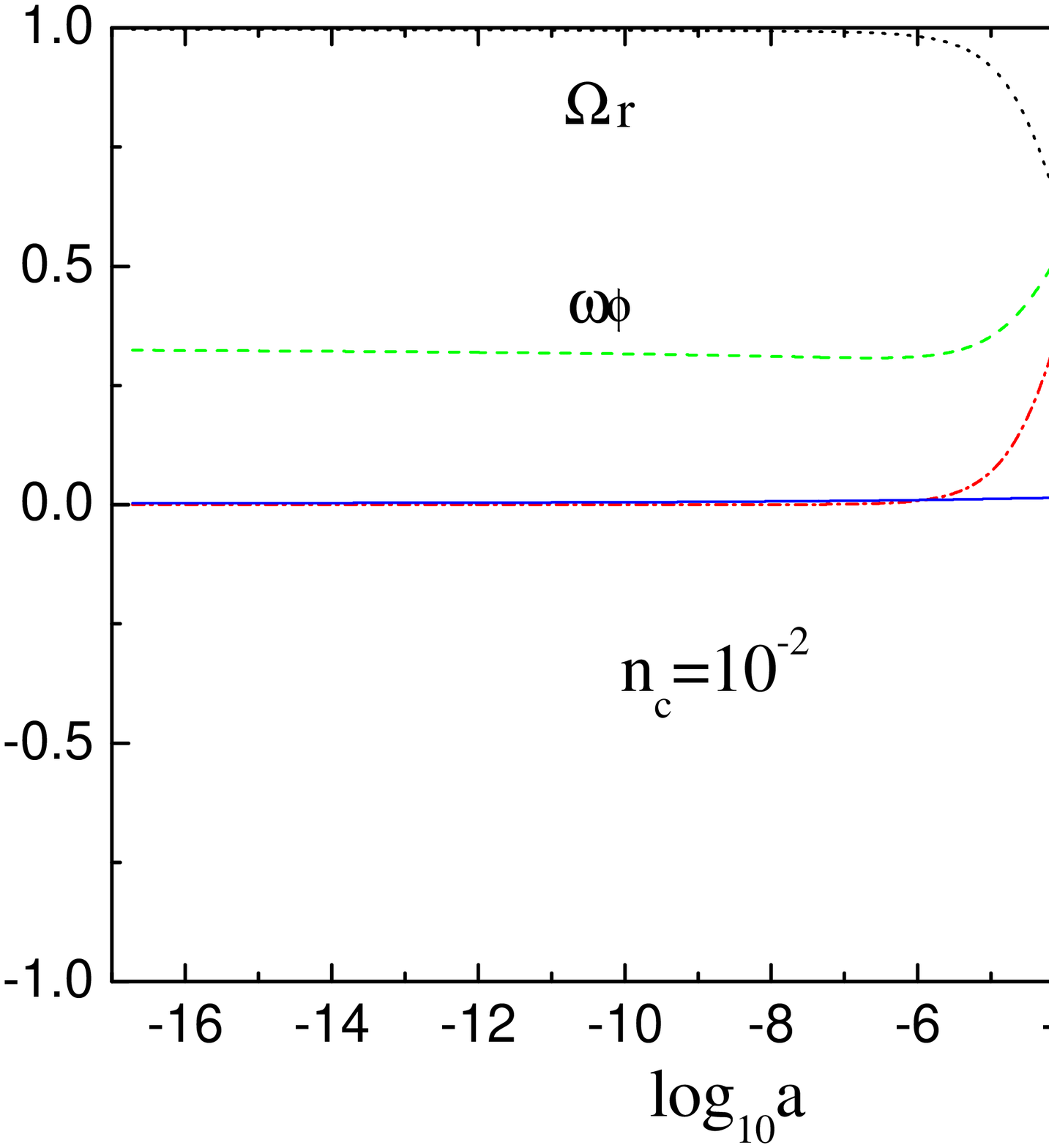, width=8cm}
\psfig{file=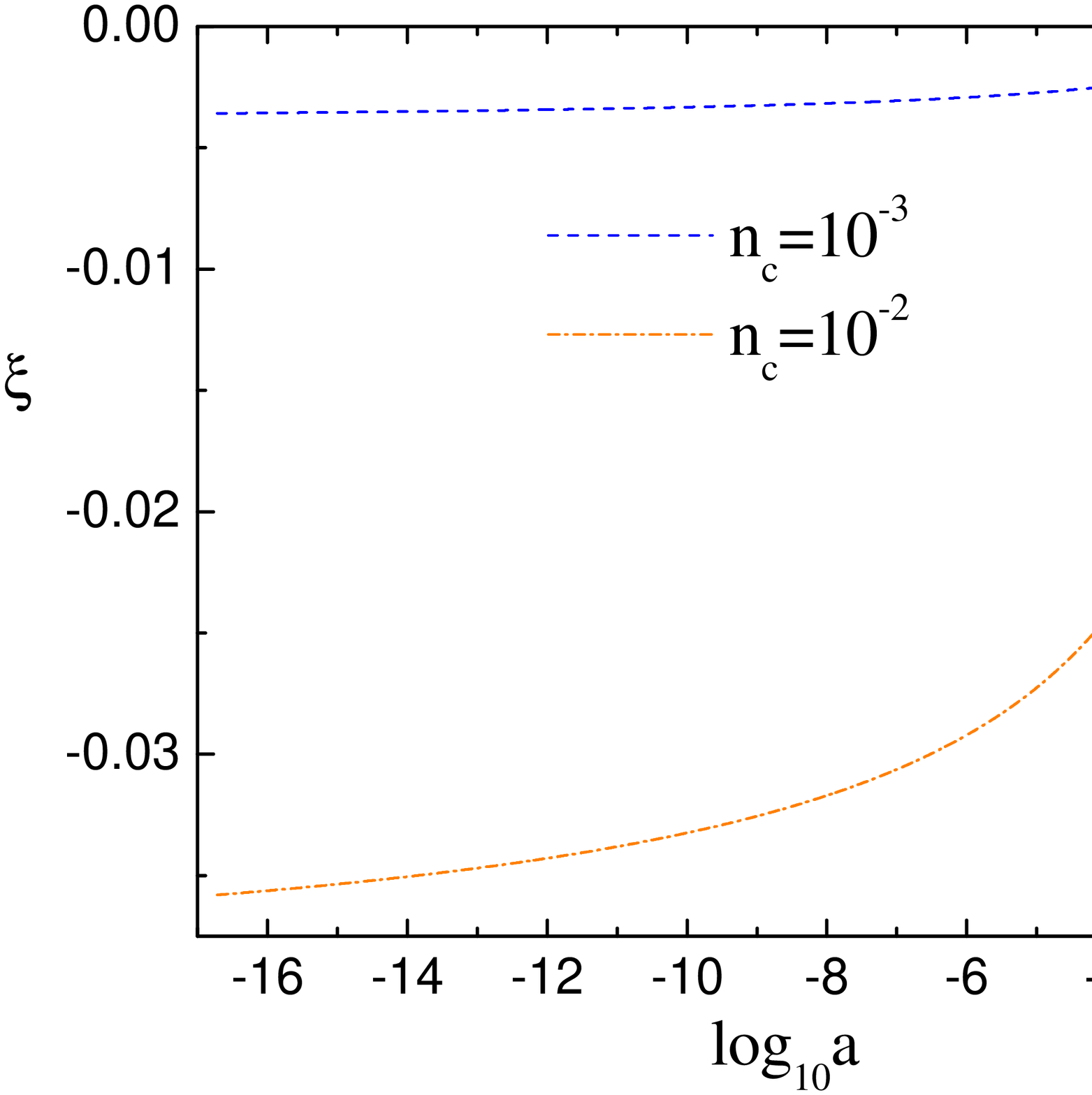, width=8cm} } \vspace{-2cm} \caption{
(a) Cosmological evolution
of the equation of state parameter, $\omega_{\phi}$, and the
energy density parameters, $\Omega_{i}$, of each component for
$\lambda =5$ and $n_c=10^{-2}$.
The evolution of the parameters is similar for other
choices of $\lambda$. (b) Deviations of the scaling of the CDM
energy density ($\xi$) from $a^{-3}$ for couplings
 with
$n_{c} = 10^{-3}$ (dashed line) and $n_{c} = 10^{-2}$ (dash-dotted
line). } \label{fig:Qnxi}
\end{figure}
\end{center}

From these we have the evolution of the background quantities as
shown in
 the first panel of Fig.~\ref{fig:Qnxi}.
The parameters we use in
 this figure for the
present energy density contrasts of the quintessence and
the CDM are $\Omega_{\phi}^{(0)} = 0.76$ and $\Omega_{c}^{(0)} =
0.20$, respectively. To be compatible with observational data, the
energy density of the quintessence must be subdominant during the
big-bang
 nucleosynthesis (BBN) $\Omega_{\phi}^{\rm{BBN}}(a
\sim 10^{-10}) < 0.2$ at $T \sim 1\,{\rm MeV}$~\cite{Ratra},
and the energy
 density at present must be comparable to
the preferred range for
 dark energy,
$0.60 \leq \Omega_{\phi}^{(0)}(a=0) \leq 0.85$
~\cite{CMB}.
The stronger bound on the energy density of
 the quintessence during BBN,
$\Omega_{\phi}^{\rm{BBN}}(a \sim 10^{-10}) < 0.045$~\cite{Bean},
is also well satisfied in this model.
We note that the equation of state (EOS) of the quintessence
$\omega_{\phi}
 \equiv p_\phi/\rho_\phi \approx \omega_r = 1/3$
during the radiation dominated era as in other tracker solutions.
Instead of
 falling from the tracking value of $1/3$ towards $-1$, as in the
uncoupled case, $\omega_{\phi}$ increases towards higher values
($\leq +1$), before ultimately dropping to $-1$ at later times.
This is due to the fact that when the matter density increases,
the effect on the effective potential $V_{eff}(\phi)= V(\phi) +
\rho_{c}$ given in Eq. (\ref{phieq}) drives the faster evolution
of $\phi$. From Eq.~(\ref{rhoc'}), we find that the scaling of the
CDM energy density differs from the usual $a^{-3}$ due to the
coupling : \ba \rho_{c}(a) &=& \rho_{c}^{(0)} a^{-3} \Bigl(
\exp[B_{c}(\phi(x)) - B_{c}(\phi(0))] \Bigr) \equiv \rho_{c}^{(0)}
a^{-3 + \xi}, \label{rhoca} \\ \xi \cdot x &=& B_{c}(\phi(x)) -
B_{c}(\phi(0)), \label{xi} \ea where we set the present scale
factor as one ($a^{(0)} =1$), $x = \ln a$, and $\xi$ is the
deviation of the CDM redshift as a result of the coupling to a
scalar field. This is shown in the second panel of
Fig.~\ref{fig:Qnxi}. As the coupling is increased, the magnitude
of $\xi$ is increased. The $\xi$ depends also on the scale factor.
Consequently, the location of the turnover in the matter power
spectrum will be shifted due to the coupling as well (see below).

We can find the time-component of the source term in the
unperturbed background ($\bar{Q}_{(d) 0}$) from Eqs.~(\ref{dTd}),
(\ref{sumQ}), (\ref{rhoc'}), and (\ref{rhophi'}) which is given by
\be \bar{Q}_{(c) 0} = - \bar{Q}_{(\phi) 0} = - B_{c,\bar{\phi}}
\bar{\phi}' \bar{\rho}_{c}. \label{Q0} \ee Due to the conservation
of the total energy-momentum (\ref{dTbeta}), we can find the
constraint equation of ${\cal B}_{c}$ and ${\cal B}_{\phi}$, \be
(\rho_{c} + p_{c}) {\cal B}_{c} + (\rho_{\phi} +
p_{\phi}) {\cal B}_{\phi} = 0. \label{bconstrain} \ee 
To consider the perturbation we can decompose the scalar field as
\be \phi(\eta, \vec{x}) =  \bar{\phi}(\eta) + \delta \phi (\eta,
\vec{x}) \label{pphi}, \ee where $\bar{\phi}$ is the unperturbed
part and $\delta \phi$ is the perturbed part of the scalar field.
We will express the perturbed parts of each quantities by means of
Fourier expansions. So the above perturbed scalar field will be
expressed as \be \delta \phi (\eta, \vec{x}) = \sum_{k} \delta
\phi_{k}(\eta) e^{i\vec{k} \cdot \vec{x}}, \label{Fourier1} \ee
where \be \delta \phi_{k}(\eta) = \frac{1}{V} \int \delta
\phi(\eta, \vec{x}) e^{-i \vec{k} \cdot \vec{x}} d^3{\vec x}.
\label{Fourier2} \ee Therefore, the
energy-momentum tensor of the scalar field can be decomposed into
the unperturbed part and the perturbed one. The background
energy-momentum tensors are \ba ^{(0)}T^{0}_{(\phi)0} &=& - \Bigl(
\frac{\bar{M}^2}{2 a^2} \bar{\phi'}^{2} + V(\bar{\phi}) \Bigr)
\equiv - \bar{\rho}_{\phi},
\label{0rhophi} \\
^{(0)}T^{i}_{(\phi)j} &=& \Bigl(\frac{\bar{M}^2}{2 a^2}
\bar{\phi'}^{2} - V(\bar{\phi})\Bigr) \delta^{i}_{j} \equiv
\bar{p}_{\phi} \delta^{i}_{j} \equiv \omega_{\phi}
\bar{\rho}_{\phi} \delta^{i}_{j}, \label{0pphi} \ea and
the first-order perturbed parts are \ba \delta T^{0}_{(\phi)0}
&=& \frac{1}{a^2} \Bigl( \bar{M}^2 \bar{\phi'}^{2} \Psi -
\bar{M}^2 \bar{\phi}' \delta \phi' - a^2 \frac{\partial
V(\bar{\phi})}{\partial \bar{\phi}} \delta \phi
\Bigr), \label{deltarhophi} \\
\delta T^{i}_{(\phi)j} &=& \frac{1}{a^2} \Bigl( - \bar{M}^2
\bar{\phi'}^{2} \Psi + \bar{M}^2 \bar{\phi}' \delta \phi' - a^2
\frac{\partial V(\bar{\phi})}{\partial \bar{\phi}} \delta \phi \Bigr)
\delta^{i}_{j}, \label{deltapphi} \\
\delta T^{0}_{(\phi)i} &=& -\frac{\bar{M}^2}{a^2} \bar{\phi}'
\partial_{i} (\delta \phi). \label{delta0iphi} \ea We can repeat
the similar consideration for the other components. We will regard
the CDM as a perfect fluid of energy density $\rho_{c}$ and pressure
$p_{c}$. To the linear order in the perturbations the energy
momentum tensors are given by \ba T^{0}_{(c) 0} &=& -\Biggl(
\bar{\rho}_{c} + \frac{\partial B_{c}(\bar{\phi})} {\partial
\bar{\phi}} \delta \phi \bar{\rho}_{c} + \delta \rho_{c} \Biggr)
\equiv -\Biggl(
\bar{\rho}_{c} + \delta \bar{\rho}_{c} \Biggr), \label{rhoc} \\
T^{0}_{(c) i} &=& ( \bar{\rho}_{c} + \bar{p}_{c})
{\it v}_{(c) i} = -T^{i}_{(c) 0}, \label{T0ic} \\
T^{i}_{(c) j} &=& \Biggl( \bar{p}_{c} + \frac{\partial
B_{c}(\bar{\phi})} {\partial \bar{\phi}} \delta \phi \bar{p}_{c} +
\delta p_{c} \Biggr) \delta^i_j + \Sigma^i_{(c) j} \equiv \Biggl(
\bar{p}_{c} + \delta \bar{p}_{c} \Biggr) \delta^i_j +
\Sigma^i_{(c) j}, \label{pc} \ea where ${\it v}^{i} =
dx^{i}/d\eta$ and $\Sigma^i_{j}$ is an anisotropic shear
perturbation. Here we define the perturbed part of the CDM as
$\delta \bar{\rho}_{c} = B_{c,\bar{\phi}} \delta \phi \bar{\rho}_{c} +
\delta \rho_{c}$, where the first term of the right hand side
is due to the coupling
 of the scalar field to  t
the CDM~\footnote{We can show this as follows: $\rho_{c} =
e^{B_c(\phi)} \rho_{c}^{*} = ( e^{B_c(\bar{\phi})} +
B_c(\bar{\phi})_{,\bar{\phi}} \delta \phi e^{B_c(\bar{\phi})})
(\bar{\rho}_{c}^{*} + \delta \rho_{c}^{*}) \sim \bar{\rho}_{c} +
B_c(\bar{\phi})_{,\bar{\phi}} \delta \phi \bar{\rho}_{c} + \delta
\rho_{c}$, where we means $B_c(\bar{\phi})_{,\bar{\phi}} =
\partial B_c(\bar{\phi})/ \partial \bar{\phi}$ and $\rho_{c}^{*}$ is the bare
energy density of the CDM.}. For the baryons and the radiation we
have the following equations, \ba T^{0}_{(\gamma) 0} &=& -\Bigl(
\bar{\rho}_{\gamma} + \delta \rho_{\gamma} \Bigr), \label{rhogamma} \\
T^{0}_{(\gamma) i} &=& \Biggl( \bar{\rho}_{\gamma} +
\bar{p}_{\gamma} \Biggr) {\it v}_{(\gamma)i} = - T^{i}_{(\gamma) 0},
 \label{T0igamma} \\
T^{i}_{(\gamma) j} &=& \Bigl( \bar{p}_{\gamma} + \delta p_{\gamma}
\Bigr) \delta^i_j + \Sigma^i_{(\gamma) j}. \label{pgamma} \ea For a
flat Friedmann-Robertson-Walker universe we can collect the
unperturbed equations for each species: \ba 3 {\cal H}^2 &=&
\frac{a^2}{\bar{M}^2} \Bigl( \bar{\rho}_{r} + \bar{\rho}_{b} +
\bar{\rho}_{c} + \bar{\rho}_{\phi} \Bigr) \equiv
\frac{a^2}{\bar{M}^2} \sum_{\beta} \bar{\rho}_{\beta} \equiv
\frac{a^2}{\bar{M}^2} \bar{\rho}_{cr}, \label{H} \\
{\cal H}' &=& - \frac{a^2}{6 \bar{M}^2} \sum_{\beta} (1 + 3
\omega_{\beta}) \bar{\rho}_{\beta} = - \sum_{\beta} \frac{(1 + 3
\omega_{\beta})}{2} \bar{\Omega}_{\beta} {\cal H}^2, \label{H'} \\
\bar{\rho}_{\gamma}' &=& - 3 {\cal H} (\bar{\rho}_{\gamma} + \bar{p}_{\gamma})
,
\label{barrhogamma'} \\
\bar{\rho}_{c}' &=& - 3 {\cal H} (\bar{\rho}_{c} + \bar{p}_{c}) +
B_{c,\bar{\phi}} \bar{\phi}' \bar{\rho}_{c}
\equiv -3 {\cal H}(\bar{\rho}_{c} + \bar{p}_{c})(1 - \bar{{\cal B}}_{c}), \label{barrhoc'} \\
\bar{\rho}_{\phi}' &=& -3 {\cal H}(\bar{\rho}_{\phi} +
\bar{p}_{\phi}) - B_{c,\bar{\phi}} \bar{\phi}' \bar{\rho}_{c}
\equiv -3 {\cal H}(\bar{\rho}_{\phi} + \bar{p}_{\phi})(1 -
\bar{{\cal B}_{\phi}}). \label{barrhophi'}\ea If we include terms
up to the first order of the energy transfer vector (\ref{dTd}),
then we can write $Q_{(d) \nu}$ as \ba Q_{(d) 0} &=& -
\bar{Q}_{(d)} (1 + \Psi) - \delta Q_{(d)}, \label{Q0p} \\ Q_{(d)
i} &=& \Bigl( f_{(d)} + \bar{Q}_{(d)} v_{(d) i} \Bigr)_{,i}.
\label{Qip} \ea where $\delta Q_{(d)}$ and $f_{(d)}$ is the energy
and momentum transfer of the CDM or quintessence, respectively.
These terms should be included in the coupled quintessence models
when we use the conformal Newtonian gauge \cite{Kodama}. If we
missed this term, then we would have $2 \Psi B_{c,\bar{\phi}}$
instead of $3 \Psi B_{c,\bar{\phi}}$ in the last term of Eq.
~(\ref{deltaphi}). With using Eqs.~(\ref{Q0p}) and (\ref{Qip}), we
can find the perturbed part of the energy momentum conservation
equations in $k$-space. The perturbed equations for the scalar
field and the fluids which can be obtained from
Eqs.~(\ref{dTgamma}) and (\ref{dTd}) are given by \ba \delta \phi
'' &+& k^2 \delta \phi + 2 {\cal H} \delta \phi' + 2 \Psi
\frac{a^2}{\bar{M}^2} V_{,\bar{\phi}} + \frac{a^2}{\bar{M}^2}
V_{,\bar{\phi}\bar{\phi}} \delta \phi - (\Psi' + 3 \Phi')
\bar{\phi}' \nonumber
\\ &=& - \frac{a^2}{\bar{M}^2} \bar{\rho}_{c} \Biggl(
B_{c,\bar{\phi}\bar{\phi}} \delta \phi + B_{c,\bar{\phi}}^2
\delta \phi + B_{c,\bar{\phi}} \delta_{c} + 3 \Psi
B_{c,\bar{\phi}} \Biggr) \nonumber
\\ &=& - \frac{a^2}{\bar{M}^2} \bar{\rho}_{c} \Biggl(
B_{c,\bar{\phi}\bar{\phi}} \delta \phi + B_{c,\bar{\phi}}
\bar{\delta}_{c} + 3 \Psi B_{c,\bar{\phi}} \Biggr)
,
\label{deltaphi} \\
\frac{\delta \rho_{c}'}{\bar{\rho}_{c}} &=& -(1 + \omega_{c})
(\theta_{c} - 3 \Phi') - 3 {\cal H} \Bigl(\frac{\delta
p_{c}}{\delta \rho_{c}} + 1 \Bigr) \delta_{c} + B_{c,\bar{\phi}}
\bar{\phi}' \Bigl( \delta_{c} + \Psi \Bigr), \label{deltarhoc} \\
\frac{\delta \bar{\rho}_{c}'}{\bar{\rho}_{c}} &=& -(1 +
\omega_{c}) (\theta_{c} - 3 \Phi') - 3 {\cal H} \Bigl(\frac{\delta
\bar{p}_{c}}{\delta \bar{\rho}_{c}} + 1 \Bigr) \bar{\delta}_{c} +
B_{c,\bar{\phi} \bar{\phi}} \bar{\phi}' \delta \phi \nonumber
\\ &+& B_{c,\bar{\phi}} \Bigl(
\bar{\phi}' \bar{\delta}_{c} + \delta \phi' + \bar{\phi}' \Psi
\Bigr), \label{deltabarrhoc} \\
\delta_{c}' &=& -(1 + \omega_{c})(\theta_{c} - 3 \Phi') - 3 {\cal
H} \Bigl(\frac{\delta p_{c}}{\delta \rho_{c}}
- \omega_{c} \Bigr) \delta_{c} + B_{c,\bar{\phi}} \bar{\phi}' \Psi, \label{deltac} \\
\bar{\delta}_{c}' &=& -(1 + \omega_{c})(\theta_{c} - 3 \Phi') - 3
{\cal H} \Bigl(\frac{\delta p_{c}}{\delta \rho_{c}} - \omega_{c}
\Bigr) \bar{\delta}_{c} + B_{c,\bar{\phi} \bar{\phi}}
\bar{\phi}' \delta \phi + B_{c,\bar{\phi}} \delta \phi'
\nonumber \\ &+& B_{c,\bar{\phi}} \bar{\phi}' \Psi, \label{bardeltac} \\
\theta_{c}' &=& - {\cal H} (1 -3 \omega_{c}) \theta_{c} -
\frac{\omega_{c}'}{1 + \omega_{c}} \theta_{c} + \frac{\delta
p_{c}/\delta \rho_{c}}{ 1 + \omega_{c}} k^2 \delta_{c} - k^2
\sigma_{c} + k^2 \Psi \nonumber \\ &+&
\fr{\omega_{c}}{(1+\omega_{c})} k^2 B_{c,\bar{\phi}} \delta \phi
- \fr{\omega_{c}}{(1+\omega_{c})} B_{c,\bar{\phi}} \bar{\phi}'
\theta_{c} \label{thetac} \\
&=& - {\cal H} (1 -3 \omega_{c}) \theta_{c} - \frac{\omega_{c}'}{1
+ \omega_{c}} \theta_{c} + \frac{\delta \bar{p}_{c}/\delta
\bar{\rho}_{c}}{ 1 + \omega_{c}} k^2
\bar{\delta}_{c} - k^2 \sigma_{c} + k^2 \Psi \nonumber \\
&-& \fr{\omega_{c}}{(1+\omega_{c})} B_{c,\bar{\phi}} \bar{\phi}'
\theta_{c}, \label{barthetac} \ea where we define $\delta_{c} =
\delta \rho_{c}/ \bar{\rho}_{c}$, $\bar{\delta}_{c} = \delta
\bar{\rho}_{c}/ \bar{\rho}_{c}$, $\theta_c=i \vec{k} \cdot
\vec{v}$, $\omega_c$ denotes the equation of state parameter (EOS)
of the CDM, and $\sigma_c$ is related to the CDM anisotropic
stress perturbation $\Pi_{c}$, by $\sigma_c = 2 \Pi_{c} \,
\bar{p_{c}}/3(\bar{\rho}_c + \bar{p}_c)$. We express both
$\delta_{c}'$ and $\bar{\delta}_{c}'$ explicitly in order to show
that the equations of the energy density perturbation can be
different with different definitions. However, it is the coupled
energy density which is measured in observations. As such, we will
use $\bar{\delta}_{\beta}$ as the energy density contrast of each
species. 
Every term containing $B_{c}(\phi)$ in the above equations comes
from the coupling. If we drop all these terms, then they are
obviously identical to the expressions Ref.~\cite{MaB}. If we have
more than one ideal gas components, then we have entropy
perturbation ($\delta S$) which can be expressed as  \be \delta
p_{\tot} = \sum_{\beta} \Bigl[ (\partial p_{\beta}/\partial
\rho_{\beta})|_{S} \, \delta \rho_{\tot} + (\partial p_{\beta}/
\partial S_{\beta})|_{\rho} \, \delta S_{\tot} \Bigr] \equiv c_{\tot}^2 \delta \rho_{\tot} +
p_{\tot} \Gamma_{\tot} , \label{deltaS} \ee where $c_{\tot}^2$ is
the overall adiabatic sound speed squared and $\Gamma_{\tot}$ is
the total entropy perturbation. We will consider this more
carefully in the following section. We can write the perturbed
equations for the CDM from the above generic perturbation
equations (\ref{bardeltac}) and (\ref{barthetac}). However due to
the coupling between the baryons and the photons we also need to
consider the Thomson scattering term in the baryon-photon fluid.
After including this we have the following equations.
\ba \delta_{b}' &=& -\theta_{b} + 3 \Phi', \label{deltab} \\
\theta_{b}' &=& -{\cal H} \theta_{b} + c_{s}^2 k^2 \delta_{b} +
k^2 \Psi + \frac{4 \bar{\rho}_{\gamma}}{3 \bar{\rho}_{b}} a n_{e}
\sigma_{T} (\theta_{\gamma} - \theta_{b}), \label{thetab} \\
\delta_{c}' &=& -\theta_{c} + 3 \Phi' + B_{c,\bar{\phi}} \bar{\phi}' \Psi, \label{deltac2} \\
\bar{\delta}_{c}' &=& -\theta_{c} + 3 \Phi' + B_{c,\bar{\phi}
\bar{\phi}} \bar{\phi}' \delta \phi +
B_{c,\bar{\phi}} \delta \phi' + B_{c,\bar{\phi}} \bar{\phi}' \Psi, \label{bardeltac2} \\
\theta_{c}' &=& -{\cal H} \theta_{c} + k^2 \Psi, \label{thetac2} \\
\delta_{r}' &=& - \frac{4}{3} \theta_{r} + 4 \Phi'
,
\label{deltar} \\
\theta_{r}' &=& k^2 \Bigl( \frac{1}{4} \delta_{r} - \sigma_{r}
\Bigr) + k^2 \Psi + a n_{e} \sigma_{T} (\theta_{b} -
\theta_{\gamma}), \label{thetar} \ea where $n_{e}$ is the electron
number density, $\sigma_{T}$ is the cross section for the Thomson
scattering.

\section{Entropy perturbation}
\setcounter{equation}{0}


Let us start from the definition of the total energy density
perturbation and that of the total pressure perturbation, \ba
\delta \bar{\rho}_{\rm{tot}} &=& \sum_{\beta} \delta
\bar{\rho}_{\beta}, \label{deltarhotot} \\
\delta \bar{p}_{\rm{tot}} &=& \sum_{\beta} \delta \bar{p}_{\beta}
. \label{deltaptot} \ea For a given $p_{\rm{tot}}(\rho,S)$, the
pressure fluctuation can be expressed as \be \delta p_{\tot}
\equiv c_{\tot}^2 \delta \rho_{\tot} + p_{\tot} \Gamma_{\rm{int}}
+ p_{\tot} \Gamma_{\rel}, \label{deltaptot2} \ee where $S$ is an
entropy, $c_{\tot}^2$ is the overall adiabatic sound speed
squared, $\Gamma_{\rm{int}}$ and $\Gamma_{\rel}$ are the intrinsic
and the relative entropy perturbations respectively. We have \ba
p_{\tot} \Gamma_{\rm{int}} &=& \sum_{\beta} p_{\beta}
\Gamma_{\beta} ,
\label{Gammaint} \\
p_{\tot} \Gamma_{\rel} &=& \sum_{\beta} (c_{\beta}^2 - c_{\tot}^2)
\delta \rho_{\beta}, \label{Gammarel} \\
c_{\tot}^2 &=& \frac{\sum_{\beta}
c_{\beta}^2\rho_{\beta}'}{\rho_{\tot}'}, \label{ctot} \ea where
$\Gamma_{\rm{int}}$ is the sum of the intrinsic entropy
perturbation of each fluid and $\Gamma_{\rel}$ arises from the
relative evolution between fluids with different sound speeds. As
we mentioned before the energy momentum of each species may not be
conserved due to the scalar field coupling even though the total
energy momentum does conserve. We can rewrite the equation for the
total adiabatic sound speed (\ref{ctot}) as \be c_{\tot}^2 =
\sum_{\beta} c_{\beta}^2 (1 -{\cal B}_{\beta}) \frac{\rho_{\beta}
+ p_{\beta}}{\rho_{\tot} + p_{\tot}}. \label{ctot2} \ee Now we can
rewrite the relative entropy perturbation as \be p_{\tot}
\Gamma_{\rel} = \frac{1}{2} \sum_{\beta,\alpha}
\frac{(\rho_{\beta} + p_{\beta})(\rho_{\alpha} + p_{\alpha})}
{\rho_{\tot} + p_{\tot}}(c_{\beta}^2 - c_{\alpha}^2) S_{\beta
\alpha} + \sum_{\beta} {\cal B}_{\beta} c_{\beta}^2 (\rho_{\beta}
+ p_{\beta}) \Delta, \label{Gammarel2} \ee where \ba S_{\beta
\alpha} &=& \Delta_{\beta} - \Delta_{\alpha}, \label{S}
\\ \Delta_{\beta} &=& \frac{\delta \rho_{\beta}}{\rho_{\beta} +
p_{\beta}}, \label{Deltabeta} \\
\Delta &=& \frac{\delta \rho_{\tot}}{\rho_{\tot} + p_{\tot}} ,
\label{Delta}  \ea where $S_{\beta \alpha}$ is the entropy
perturbation \cite{Kodama}. Due to the $\Delta$-term the relative
entropy perturbation is non-vanishing even without the
non-adiabatic perturbation. This is improper, so we redefine the
new quantities as \ba \hat{\Delta}_{\beta} &=& \frac{\delta
\rho_{\beta}}{(1 - {\cal
B}_{\beta})(\rho_{\beta} + p_{\beta})}, \label{hatDeltabeta} \\
\hat{S}_{\beta \alpha} &=& \hat{\Delta}_{\beta} -
\hat{\Delta}_{\alpha}. \label{S2} \ea With these quantities we can
rewrite Eq.~(\ref{Gammarel2}) as
\be p_{\tot} \Gamma_{\rel} = \frac{1}{2} \sum_{\beta,\alpha}
\frac{(1 - {\cal B}_{\beta})(1 - {\cal B}_{\alpha})(\rho_{\beta} +
p_{\beta})(\rho_{\alpha} + p_{\alpha})} {\rho_{\tot} +
p_{\tot}}(c_{\beta}^2 - c_{\alpha}^2) \hat{S}_{\beta \alpha}.
\label{Gammarel3} \ee

\begin{center}
\begin{figure}
\vspace{1cm} \centerline{\psfig{file=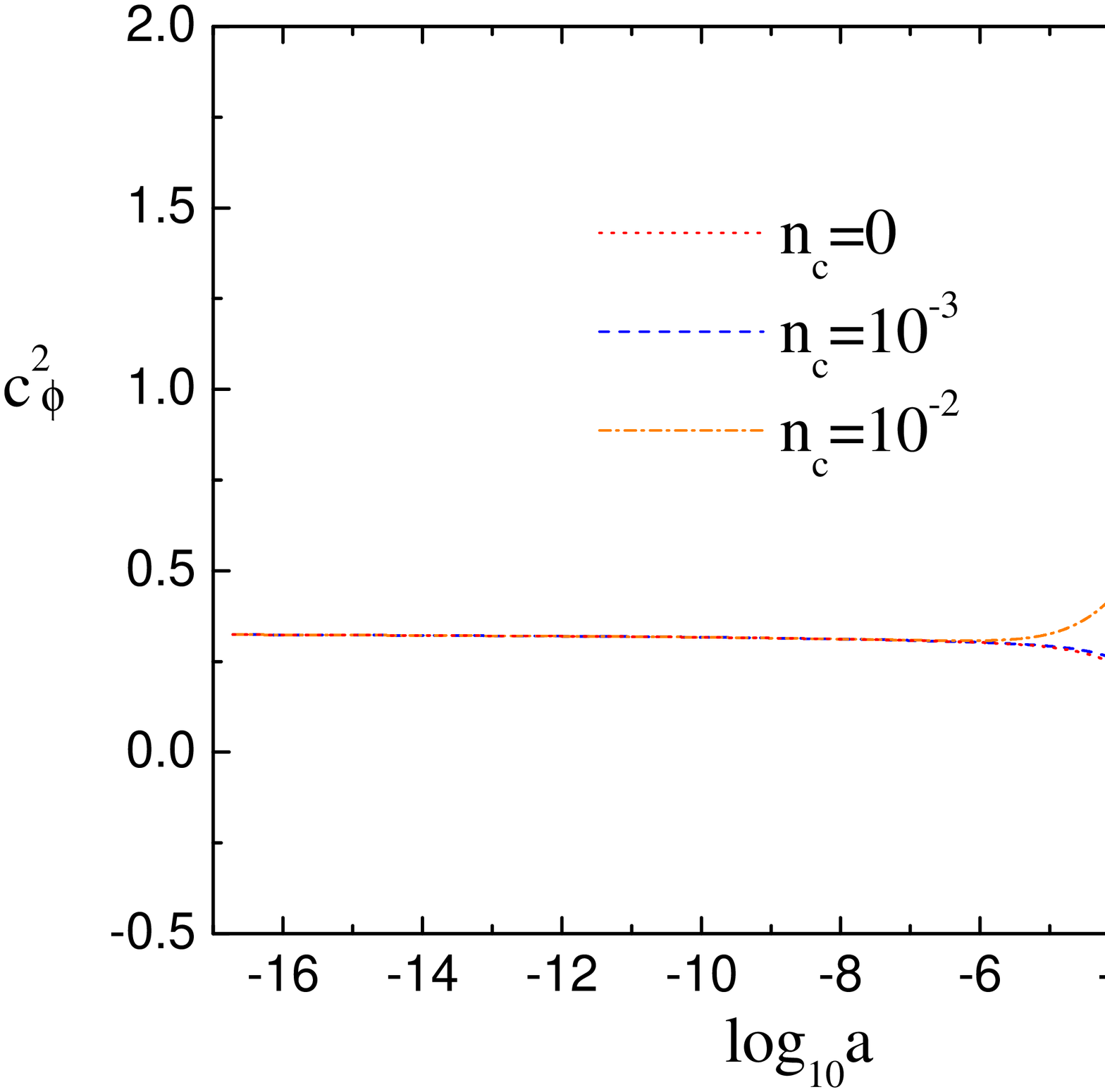, width=8cm}
\psfig{file=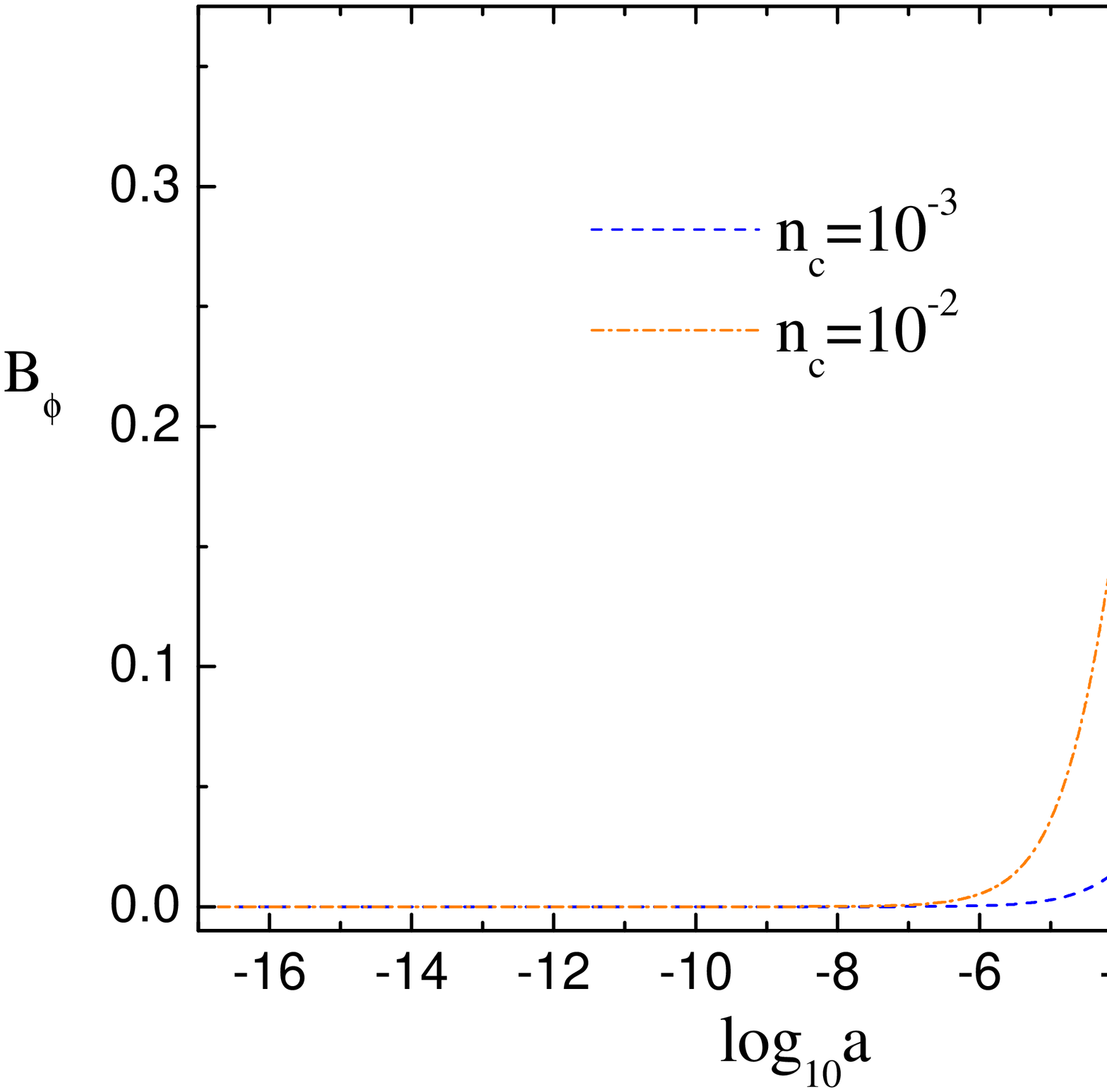, width=8cm} } \vspace{-2cm} \caption{(a)
Cosmological evolution of the adiabatic sound speed $c_{\phi}^2$
of the scalar field for different values of $n_{c} = 0$ (dotted
line), $n_{c} = 10^{-3}$ (dashed line), and $n_{c} = 10^{-2}$
(dash-dotted line) when $\lambda =5$. (b) Behavior of the coupling
modification, ${\cal B}_{\phi}$ as a function of the scale factor
$a$ for different values of $n_c$.} \label{fig:cpBn}
\end{figure}
\end{center}
\subsection{Isocurvature condition}


Due to the out of thermal equilibrium nature of the quintessence,
we need to check the isocurvature evolution of the scalar field
perturbation \cite{Isocur}. We can analytically show this in the
tracking region. First of all the adiabatic sound speed squared
$c_{\phi}^2$ of the quintessence can be represented as  \be
c_{\phi}^2 = \frac{\bar{p}_{\phi}'}{\bar{\rho_{\phi}}'} = 1 +
\frac{2 \bar{\phi}' V_{,\bar{\phi}}}{3 {\cal H} (\bar{\rho}_{\phi}
+ \bar{p}_{\phi})( 1 - {\cal B}_{\phi})} = \omega_{\phi} -
\fr{\omega_{\phi}'}{3 {\cal H} ( 1 + \omega_{\phi})(1 - {\cal
B}_{\phi})}, \label{cphi} \ee where we have used
Eq.~(\ref{rhophi'}). From this equation we can find the second
derivative of the potential, \be \frac{a^2}{\bar{M}^2}
V_{,\bar{\phi} \bar{\phi}} = \frac{3}{2} {\cal H} c_{\phi}^{2'}(1
- {\cal B}_{\phi}) + \frac{3}{2} {\cal H}^2(c_{\phi}^2 - 1)(1 -
{\cal B}_{\phi}) \Bigl[ \frac{{\cal H}'}{{\cal H}^2} - 1 -
\frac{3}{2}(c_{\phi}^2 + 1)(1 - {\cal B}_{\phi}) \Bigr] -
\frac{3}{2} {\cal H}(c_{\phi}^2 - 1) {\cal B}_{\phi}'.
\label{Vdouble} \ee The relation between ${\cal H}^2$ and ${\cal
H}'$ can be found from Eq.~(\ref{H'}) : \ba \fr{{\cal
H}'}{{\cal H}^2} &=& - \fr{1}{2} ( 1 + 3 \omega_{\tot}), \label{HH} \\
\omega_{\tot} &=& \sum_{\beta} \omega_{\beta} \Omega_{\beta},
\label{omegatot} \ea where $\omega_{\tot}$ is the weighted EOS. As
long as we use the background as shown in the pervious model, the
tracking region is well established during the radiation dominated
era~\cite{LOP}. During this era we can find the following
relation, \be c_{\phi}^2 =
\frac{\bar{p}_{\phi}'}{\bar{\rho}_{\phi}'} = \omega_{\phi} =
\omega_{r}. \label{trackingcphi} \ee This is shown in the first
panel of Fig.~\ref{fig:cpBn}. As the coupling is increased we can
have the non-monotonic behavior of $c_{\phi}^2$ as shown in the
$n_{c} = 10^{-2}$ case. In addition, we can rewrite the coupling
term as \be {\cal B}_{\phi} = - \frac{ B_{c,\bar{\phi}}
\bar{\phi}' \bar{\rho}_{c}}{3 {\cal H} (\bar{\rho}_{\phi} +
\bar{p}_{\phi})} = - \frac{{\cal H}}{\bar{\phi}'} B_{c,\bar{\phi}}
\bar{\Omega}_{c} . \label{calBphi} \ee As shown in the second
panel of Fig.~\ref{fig:cpBn}, this term is negligible during the
radiation dominated epoch. The coupling drives a faster evolution
of $\phi$ when matter energy dominates the Universe and the
magnitude of ${\cal B}_{\phi}$ depends on the energy density of
the CDM. With these facts we can have the approximate expression
of Eq.~(\ref{Vdouble}), \be \frac{a^2}{\bar{M}^2} V_{,\bar{\phi}
\bar{\phi}} \simeq \frac{3}{2} {\cal H}^2(c_{\phi}^2 - 1) \Bigl[
\frac{{\cal H}'}{{\cal H}^2} - \frac{3}{2}(c_{\phi}^2 + 1) \Bigr]
= - \frac{3}{4} {\cal H}^2(c_{\phi}^2 - 1) \Bigl[ 3 \omega_{\tot}
+ 3 c_{\phi}^2 + 4 \Bigr], \label{Vdouble2} \ee where we have used
Eq.~(\ref{HH}) in the second equality. The isocurvature mode in
the radiation dominated epoch can be obtained from
Eq.~(\ref{deltaphi}) and using the fact that $a \propto \eta$
during the radiation dominated era. To check this we can put $\Phi
= 0$ and then we have \be \delta \phi'' + \frac{4}{(3 \omega_{r} +
1) \eta} \delta \phi' + \Bigl[ k^2 - (c_{\phi}^2 - 1) \frac{3}{(3
\omega_{r} + 1)^2 \eta^2} (3 \omega_{r} + 3 c_{\phi}^2 + 4 )
\Bigr] \delta \phi = 0. \label{deltaphi2} \ee If we use the
tracking solution (\ref{trackingcphi}), then we can rewrite the
above equation as \be \delta \phi'' + \frac{2}{\eta} \delta \phi'
+ \Bigl[ k^2 + \frac{3}{\eta^2} \Bigr] \delta \phi = 0.
\label{deltaphi3} \ee The solutions of this equation are Bessel
functions, \be \delta \phi(\eta) = {\rm const} ~ \eta^{-1/2} ~
J_{\pm |i \sqrt{11}/2|} (k \eta) . \label{soldeltaphi1} \ee Thus
both solutions decay in time. At the superhorizon scale the
$k$-dependent term can be neglected and  we obtain the power-law
solutions, $\delta \phi \propto
 \eta^{\nu}$,
where the power index $\nu = (-1 \pm i \sqrt{11})/2$.
Therefore, any initial nonzero isocurvature fluctuation of the
quintessence is damped to zero with time. We will use
only the adiabatic perturbations in the following section.
\begin{center}
\begin{figure}
\vspace{1cm} \centerline{\psfig{file=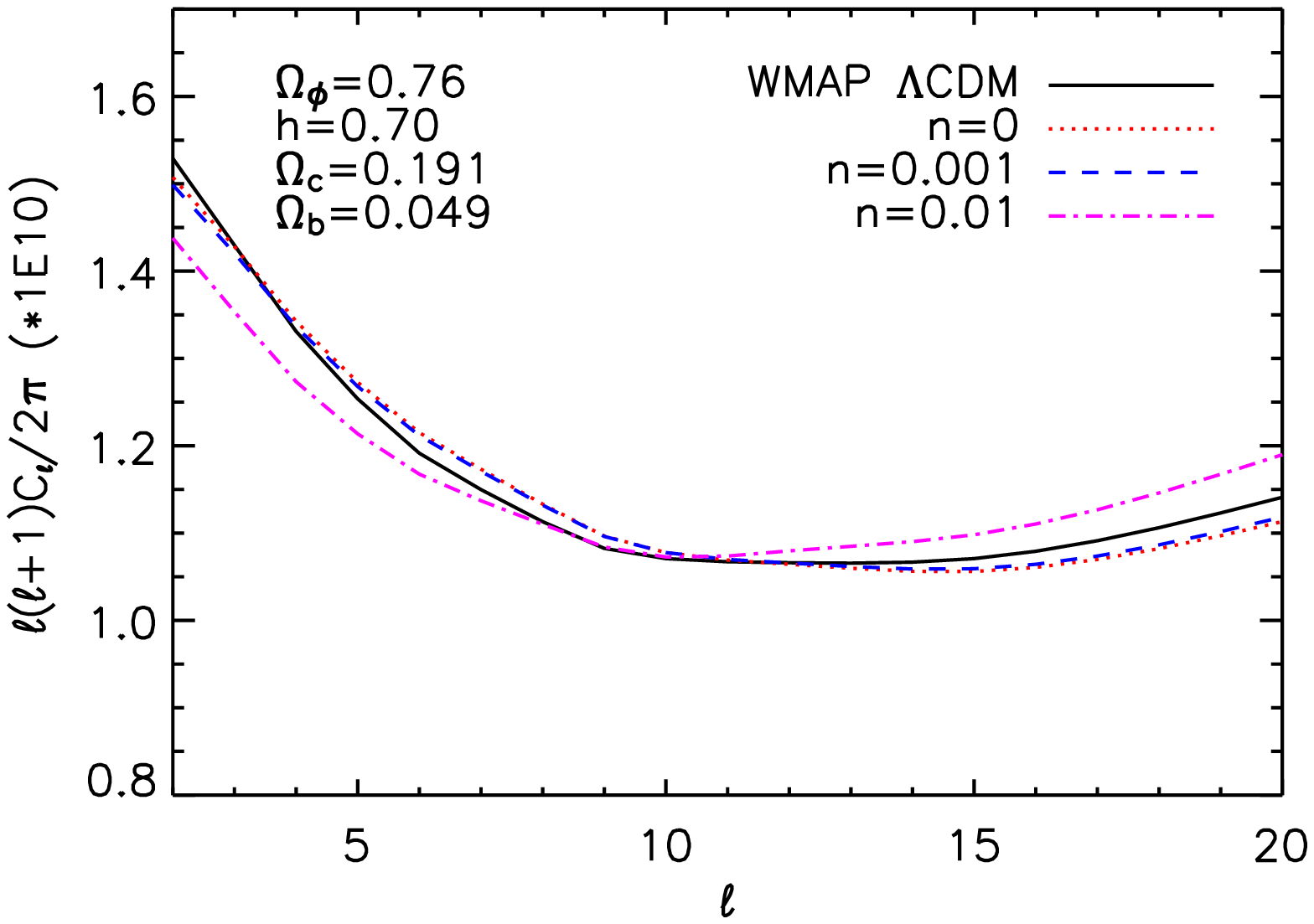,
width=8cm} \psfig{file=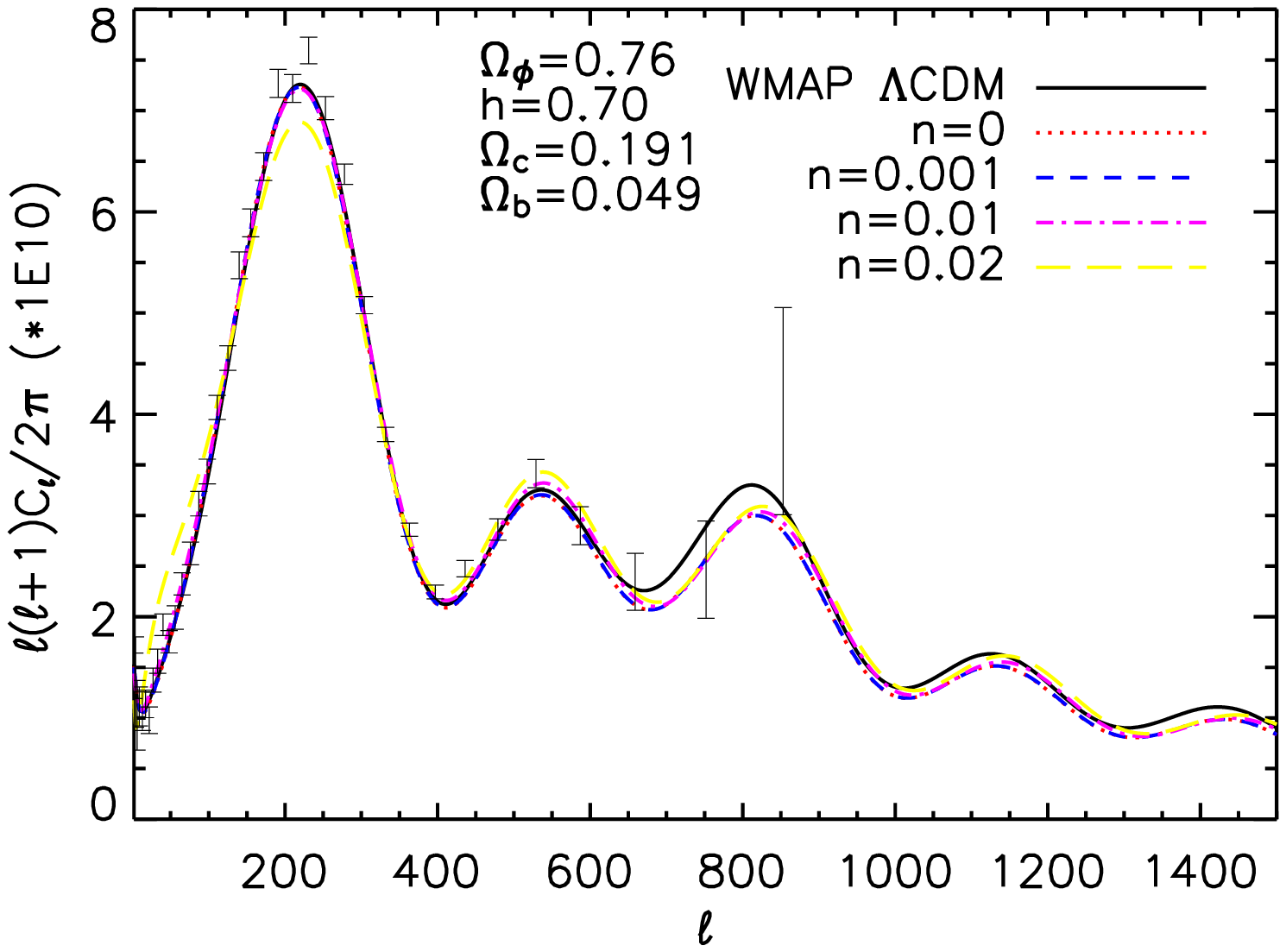,
width=8cm} } \vspace{0cm} \caption{ (a) CMB large-scale anisotropy power
spectra of $\Lambda$CDM (solid line), minimally coupled $n_{c} =
0$ (dotted line), and non-minimally coupled $n_{c} = 10^{-3},
10^{-2}$ (dashed, dash-dotted line respectively) quintessence
models. (b) Same spectra for the entire scales.}
\label{fig:cl}
\end{figure}
\end{center}
\section{Effects of coupling}
\setcounter{equation}{0}
The non-minimally coupled quintessence models have been
investigated as a possible solution for the late time coincidence
problem \cite{coupQ}. The coupling gives rise to the additional
mass and source terms of the evolution equations for CDM and scalar field
perturbations. This also affects the perturbation of radiation
indirectly through the background bulk ${\cal H}$ and the metric
perturbations \cite{Bean2}. The value of the energy density
contrast of the CDM ($\Omega_{c}$) is increased in the past when the
coupling is increased.

\subsection{CMB} \setcounter{equation}{0}

The temperature anisotropy measured in a given direction of the
sky can be expanded in spherical harmonics as \be \Theta \equiv
\frac{\Delta T}{T}(\hat{n})=\sum_{\ell m}a_{\ell m}Y_{\ell m}
(\hat{n}), \label{Theta} \ee where $\Theta$ is the temperature
brightness function that is the fractional perturbation of the
temperature of the photons $T = T_{0}(1 + \Theta)$, $\hat{n}$ is
the direction of the photon momentum, and $a_{\ell m}$ are the
multipoles. We can also expand the brightness function as a
Legendre polynomial, $P_{\ell}$ : \be \Theta = \sum_{\ell}
(-i)^{\ell} \, (2 \ell + 1) \, \Theta_{\ell} \, P_{\ell}
.
\label{Theta2} \ee From the inflationary scenario, the multipoles
are Gaussian random variables which satisfy \be \langle a^*_{\ell
m} \, a_{\ell' m'}\rangle = C_{\ell} \, \delta_{\ell \ell'} \,
\delta_{m m'}. \label{Cl} \ee The angular power spectrum
($C_{\ell}$) contains all the information about the statistical
properties of CMB and is defined as \cite{Hu}  \be \frac{2 \ell +
1}{4 \pi} C_{\ell} = \frac{1}{2 \pi^2} \int d \eta \frac{dk}{k}
\frac{k^3 |\Theta_{\ell}(k,\eta)|^2}{2 \ell +1}. \label{spectrum}
\ee In the standard recombination model the acoustic oscillations
will be frozen into the CMB. A generalization of the
free-streaming equation in a flat universe gives the resulting
anisotropies : \be \fr{\Theta_{\ell}(\eta)}{2 \ell + 1} =
[\Theta_0 + \Psi](\eta_{ls}) j_{\ell}(k(\eta - \eta_{ls})) +
\Theta_{1}(\eta_{ls}) \fr{1}{k} \fr{d}{d \eta} j_{\ell}(k(\eta -
\eta_{ls})) + \int_{\eta_{\,ls}}^{\eta} (\Psi' - \Phi') j_{\ell}
(k(\eta - \tilde{\eta})) d \tilde{\eta} \label{Spectrum2} \ee
where $j_{\ell}$ is the spherical Bessel function and
$\eta_{\,ls}$ is the conformal time at last scattering. Photon
density perturbation is related to the temperature perturbation in
the matter rest frame : \be \delta_{r} = 4 \Theta_0 + 4 {\cal H}
\fr{\theta_{\tot}}{k^2}, \label{deltarm} \ee and the gravitational
potentials will be given in Eq. (\ref{deltaG00}) -
(\ref{deltaGij}) in the following section. We use the fact that in
the absence of anisotropic stress, the two scalar potentials
$\Psi$ and $\Phi$ defined in the conformal Newtonian gauge
(\ref{CNG}) are equal and they coincide with the usual
gravitational potential in the Newtonian limit.

Now, we investigate the effects of non-minimal coupling of a
scalar field to the CDM on the CMB power spectrum. Firstly, the
Newtonian potential at late times changes more rapidly as the
coupling increases, as shown in Eq.~(\ref{deltaG00}). This leads
to an enhanced ISW effect as indicated in the last term of
Eq.~(\ref{Spectrum2}). Thus we have a relatively larger $C_{\ell}$
at large scales ({\it i.e.} small $\ell$). Thus, if the CMB power
spectrum normalized by COBE, then we will have smaller quadrupole
\cite{COBE}. This is shown in the first panel of
Fig.~\ref{fig:cl}. One thing that should be emphasized is that we
use different parameters for the $\Lambda$CDM and the coupled
quintessence models to match the amplitude of the first CMB
anisotropy peak. The parameter used for the quintessence model is
indicated in Fig.~\ref{fig:cl} ({\it i.e.} $\Omega_{\phi}^{(0)} =
0.76$, $\Omega_{m}^{(0)} = 0.191$, $\Omega_{b}^{(0)} = 0.049$, and
$h = 0.7$, where $h$ is the present Hubble parameter in the unit
of $100 {\rm km s^{-1} Mpc^{-1}}$). However, these parameters are
well inside the $1 \, \sigma$ region given by the WMAP data. We
use the WMAP parameters for the $\Lambda$CDM model ({\it i.e.}
$\Omega_{\phi}^{(0)} = 0.73$, $\Omega_{m}^{(0)} = 0.23$,
$\Omega_{b}^{(0)} = 0.04$, and $h = 0.72$). In both models we use
the same spectral index $n_{s} =1$. The heights of the acoustic
peaks at small scales ({\it i.e} large $\ell$) can be affected by
the following two factors. One is the fact that the scaling of the
CDM energy density deviates from that of the baryon energy
density. Therefore for the given CDM and baryon energy densities
today, the energy density contrast of baryons at decoupling
($\Omega_{b}^{(ls)}$) is getting lower as the coupling is being
increased. This suppresses the amplitude of compressional (odd
number) peaks while enhancing rarefaction (even number) peaks. The
other is that for models normalized by COBE, which approximately
fixes the spectrum at $\ell \simeq 10$, the angular amplitude at
small scales is suppressed in the coupled quintessence. This is
shown in the second panel of Fig.~\ref{fig:cl}. The third peak in
this model is smaller than that in the $\Lambda$CDM model. The
WMAP data do not show the value of the third peak but quote a
compilation of other experiments \cite{Wang}. The ratio of the
amplitude between the second and the third peaks is $1.03 \pm
0.02$. In the $\Lambda$CDM model this value is $0.986$ and in our
model these values are $1.08$ and $1.11$ for without and with the
coupling equal to $n_c=0.01$ respectively. We also show the $n_{c}
= 0.02$ case in this figure. With the same parameters this case
can be ruled out from the current data. In all cases, we have used
the CMBFAST code~\cite{CMBFAST} with the modified Boltzmann
equations to compute the CMB power spectrum.

Secondly, for $\ell>200$ we can see that the locations of the
acoustic peaks are slightly shifted to smaller scales ({\it i.e.}
larger $\ell$). This can be explained as follows. The locations of
peaks and troughs can be parametrized as \be \ell_{m} = \ell_{A}
(m - \varphi_m) = \ell_{A} (m - \bar{\varphi} - \delta \varphi_{m}
), \label{lm} \ee where $\ell_{A}$ is the acoustic scale dependent
on the geometry of the Universe, $\bar{\varphi}$ is the overall
peak shift, and $\delta \varphi_{m}$ is the relative shift of the
$m$-th peak relative to the first one~\cite{Doran}. The overall
peak shift, $\bar{\varphi}$ is given by  \be \bar{\varphi} \simeq
0.267 \Bigl(\fr{r_{ls}}{0.3} \Bigr)^{0.1}, \label{varphi} \ee
where $r_{ls} = \rho_r^{(ls)} / \rho_m^{(ls)}$ is the ratio of the
energy densities of radiation to matter at last scattering. The
shift is due to both driving effects from the decay of the
gravitational potential and contributions from the Doppler shift
of the oscillating fluid. The acoustic scale $\ell_{A}$ depends on
both the sound horizon $s_{ls}$ at decoupling and the angular
diameter distance $D$ to the last scattering surface: \be \ell_{A}
= \pi \fr{D}{s_{ls}} = \pi \fr{\eta_{0} - \eta_{\,ls}}{\bar{c}_{s}
\eta_{\, ls}} , \label{la} \ee where $\bar{c}_{s}$ is the average
sound speed before last scattering : \be \bar{c}_{s} =
{\eta}_{\,ls}^{-1} \int_{0}^{\eta_{\,ls}} c_{s} d \eta
\hspace{0.2in} {\rm with} \hspace{0.2in} c_s^{-2} = 3 + \fr{9}{4}
\fr{\rho_{b}}{\rho_{r}} . \label{cs} \ee Also from the Hubble
parameter, \be \Biggl(\frac{d a}{d \eta}\Biggr)^2 = H_{0}^2
\Biggl\{ \Omega_{r}^{(0)} + \Omega_{b}^{(0)} a + \Omega_{m}^{(0)}
a^{1+ \xi} + \fr{\rho_{\phi}}{\rho_{cr}^{(0)}} a^{4} \Biggr\},
\label{da1} \ee where $H_{0}$ is the present value of the Hubble
parameter, we can find the angular diameter distance, \be
D=\eta_{0} - \eta_{ls} = H_{0}^{-1} \int_{a_{ls}}^{a^{(0)}}
\fr{da}{\sqrt{\Omega_{r}^{(0)} + \Omega_{b}^{(0)} a +
\Omega_{m}^{(0)} a^{1+ \xi} + \rho_{\phi} / \rho_{cr}^{(0)} a^4}}.
\label{da2} \ee The sound horizon is not affected by the coupling
and the effect of coupling on the angular diameter distance is
also quite small in our model. However, the overall shift and the
relative shift are affected by the coupling. As the coupling is
increased, the value of $r_{ls}$ is decreased. Hence, the
locations of the peaks are shifted to the right. However, this
shift is quite small. We show this in the second panel of
Fig.~\ref{fig:cl}. The heights and the locations of the acoustic
peaks in various models are summarized in Table~\ref{tab:1}. There
is a significant difference for the heights of the peaks of the
second and the third peaks between the models. Thus upcoming
observations continuing to focus on resolving the higher peaks may
constrain the strength of the coupling.

\begin{table}[htb]
\begin{center}
\caption{Summary of the heights and the locations of the
considered models. The heights of peaks, $A_{i}$ have been scaled
by a factor of $10^{10}$. $A_{i:j}$ is the ratio of the $i$-th
peak height to the $j$-th. The locations of peaks, $\ell_{i}$.
\label{tab:1}} \vskip .3cm
\begin{tabular}{|c|c|c|c|c|c|c|c|c|}
\hline Model & $A_{1}$ & $A_{2}$ & $A_{3}$ & $\ell_{1}$ &
$\ell_{2}$ & $\ell_{3}$ & $A_{1:2}$ & $A_{2:3}$ \\
\hline $\Lambda$CDM & $7.26$ & $3.25$ & $3.30$ & $220$ & $536$ & $812$
& $2.23$& $0.986$ \\
\hline $n_{c} = 0$ & $7.04$ & $3.23$ & $3.00$ & $217$ & $532$ &
$811$ & $2.18$& $1.08$ \\
\hline $n_{c} = 10^{-2}$ & $7.02$ & $3.36$ & $3.03$ & $218$ &
$535$ & $818$ & $2.09$& $1.11$ \\
\hline
\end{tabular}
\end{center}
\end{table}

\subsection{Matter power spectrum}
\begin{center}
\begin{figure}
\vspace{1cm} \centerline{ \psfig{file=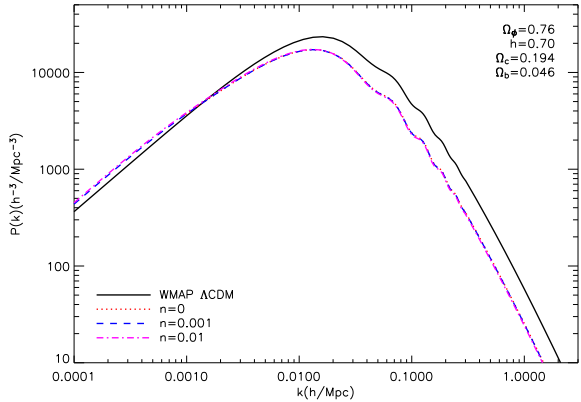,
width=8cm} } \vspace{0cm} \caption{Matter power spectra for the
models using the same parameters in Fig.~\ref{fig:cl}.}
\label{fig:mp}
\end{figure}
\end{center}
The effects of quintessence on structure formation in several
other models have been investigated \cite{WS, DSW}. Structure as a
function of physical scale size is usually described in terms of a
power spectrum : \be P(k) = \langle|\delta_{k}|^2\rangle = A
k^{n_{s}} T^{2}(k), \label{Pk} \ee where $A$ is the COBE
normalization, $n_s$ is a power index, and $T(k)$ is the transfer
function. The coupling of quintessence to the CDM can change the
shape of matter power spectrum because the location of the
turnover corresponds to the scale that entered the Hubble radius
when the Universe became matter-dominated. This shift on the scale
of matter and radiation equality is indicated in the second panel
of Fig.~\ref{fig:Qnxi}: \be a_{eq} \simeq
\fr{\rho_{r}^{(0)}}{\rho_{c}^{(0)}} \exp [B_{c}(\phi_{0}) -
B_{c}(\phi_{eq})], \label{aeq} \ee where $\rho_{r}^{(0)}$ and
$\rho_{c}^{(0)}$ are the present values of the energy densities of
radiation and CDM respectively, and the approximation comes from
the fact that the present energy density of CDM is bigger than
that of baryons ($\rho_{c}^{(0)} > \rho_{b}^{(0)}$). Increasing
the coupling shifts the epoch of matter-radiation equality further
from the present, thereby moving the turnover in the power
spectrum to smaller scale. If we define $k_{eq}$ as the wavenumber
of the mode which enters the horizon at radiation-matter equality,
then we will obtain \be k_{eq} = \fr{2 \pi}{\eta_{eq}}.
\label{keq} \ee However, from the previous subsection, we notice
that the value of $\eta_{eq}$ remains unchanged for different
couplings and this degeneracy is indicated in Fig.~\ref{fig:mp}.
We have used different parameters for the $\Lambda$CDM and
quintessence models; the matter power spectra look different
between the models. There is a slight suppression in the
quintessence models. Note that a bias factor could resolve the
discrepancy and perhaps a parameter fitting may also help this.
However, the detailed parameter fitting is beyond the scope of
this paper.

We can write the equation of the matter fluctuation in the
synchronous gauge during the matter dominated epoch : \be
\bar{\delta}_{c}'' + {\cal H} \bar{\delta}_{c}' - \fr{3}{2} {\cal
H}^2 \fr{( \delta \bar{\rho}_{c} + \delta \bar{p}_{c}
)}{\bar{\rho}_{cr}}  - \fr{(a F(\phi))'}{a} = 0, \label{deltacsyn}
\ee where the coupling term $F(\phi)$ is given by
 \be F(\phi) = B_{c,\bar{\phi}
\bar{\phi}} \bar{\phi}' \delta \phi + B_{c,\bar{\phi}} \delta
\phi' = n_c \lambda (\bar{\phi} \delta \phi)'. \label{coupc} \ee
This equation can be rewritten by the structure growth exponent
$f$, which is defined as  \be f(a) = \fr{d \ln
\bar{\delta}_{c}(a)}{d x}. \label{fa} \ee Hence, we have the
following equation for $f(a)$ which is identical to Eq.
(\ref{deltacsyn}): \be \fr{d f}{d x} + f^2 + \Bigl( 2 + \fr{d \ln
({\cal H} / a )}{d x} \Bigr) f - \fr{3}{2} \Omega_{c} = \fr{d
\Bigl[a F(\phi) \Bigr]}{d x} \fr{1}{a {\cal H} \bar{\delta}_{c}},
\label{faeq} \ee If we use Eq.~(\ref{rhoca}), then we can find
that $a \propto \eta^{2/(1-\xi)}$ and ${\cal H} = 2/(1- \xi)
\eta^{-1}$ for a matter dominated universe. As we can see from the
above Eq.~(\ref{faeq}) that the coupling term is negligible
because $\phi$ varies much slower than ${\cal H}$. Thus we can
ignore the last term in Eq.~(\ref{deltacsyn}). From this we can
rewrite the above equation in a matter dominated era : \be
\bar{\delta}_{c}'' + \fr{2}{(1-\xi) \eta} \bar{\delta}_{c}' -
\fr{6}{(1-\xi)^2 \eta^2} \bar{\delta}_{c} = 0. \label{deltacsyn2}
\ee This equation has two solutions, \be \bar{\delta}_{c}^{\pm} =
c_{\pm} \eta^{\nu_{\pm}}, \label{deltacsol} \ee where $c_{\pm}$
are arbitrary constants and \be \nu_{\pm} = \fr{-(1+\xi) \pm
\sqrt{24 + (1 + \xi)^2}}{2(1 - \xi)}. \label{nupm} \ee
$\bar{\delta}_{c}^{+}$ indicates a growing mode, which is only
relevant today because $|\bar{\delta}_{c}^{+}|$ is small at early
time and we can ignore the decaying mode $\bar{\delta}_{c}^{-}$.
If we remove the coupling effect in this solution, then we can
recover the well known solution $\delta_{c}^{+} = \eta^{2/3}$.
This effect is shown in Fig.~\ref{fig:mp}.

As the coupling is increased, we have little more matter at early
times and this increases the height of the matter power spectrum. Again
this effect is tiny and we can hardly see the difference between
various couplings.

\section{Metric perturbation}
\setcounter{equation}{0}

In addition to these, the perturbed equations of the metric can be
obtained from the Einstein equations:
\ba k^2 \Phi + 3 {\cal H} \Bigl( \Phi' + {\cal H} \Psi \Bigr) &=&
\frac{3}{2} \frac{{\cal H}^2}{\bar{\rho}_{tot}} \delta
T^0_{(tot)0}, \label{deltaG00} \\
\Phi'' + {\cal H} ( \Psi' + 2 \Phi') + \Bigl(2{\cal H}' + {\cal
H}^2 \Bigr) \Psi + \frac{k^2}{3} ( \Phi - \Psi) &=& \frac{1}{2}
\frac{{\cal H}^2}{\bar{\rho}_{tot} } \delta T^i_{(tot)i}, \label{deltaGii} \\
k^2 \Bigl( \Phi' + {\cal H} \Psi \Bigr) &=& \frac{3}{2} {\cal H}^2
\sum_{\beta} (1 + \omega_{\beta}) \bar{\Omega}_{\beta} \theta_{\beta}
,
\label{deltaG0i} \\
k^2 ( \Phi - \Psi) &=& \frac{9}{2} {\cal H}^2 \sum_{\beta} (1 +
\omega_{\beta}) \bar{\Omega}_{\beta} \sigma_{\beta}
,
\label{deltaGij} \ea
where we have used the unperturbed equations~(\ref{H}) and (\ref{H'})
.
From Eqs.~(\ref{deltaG00}) and (\ref{deltaGii}), we
can find the metric perturbation equation,
\ba \Phi'' + {\cal H}(\Psi' + 5 \Phi') + 2 ( {\cal H}' + 2{\cal
H}^2) \Psi + \frac{k^2}{3}(4\Phi - \Psi) \nonumber \\
= -\frac{a^2}{2\bar{M}^2} [ (1 - c_{tot}^2) \delta \rho_{tot} -
p_{tot} \Gamma_{int} - p_{tot} \Gamma_{rel} ]. \label{metricper}
 \ea
If we use Eq.~(\ref{deltaG00}),
this equation can be rewritten as
\ba \Phi'' + {\cal H} \Bigl[\Psi' + (2 + 3 c_{tot}^2) \Phi' \Bigr]
+ \Bigl[ 2 {\cal H}' + {\cal H}^2(1 + 3c_{tot}^2) \Bigr] \Psi +
\frac{k^2}{3} \Bigl[ (1 + 3 c_{tot}^2) \Phi - \Psi \Bigr]
\nonumber \\ = \frac{a^2}{2\bar{M}^2} \Bigl( p_{tot} \Gamma_{int}
+ p_{tot} \Gamma_{rel} \Bigr) \equiv \frac{a^2}{2\bar{M}^2}
p_{tot} \Gamma_{tot}, \label{metricper2} \ea where we define
$\Gamma_{tot} = \Gamma_{int} + \Gamma_{rel}$ in the last equality.
The last term in this equation comes from the coupling of
 the
scalar field. So even if we start from the adiabatic condition (
$p_{tot} \Gamma_{tot} =0$), we can analytically solve the above
equation for the specific case. Let us first put $\Psi = \Phi$ (no
anisotropic stress) and $k^2 c_{tot}^2 \Phi = 0$ (consider the
superhorizon scale), then Eq.~(\ref{metricper2}) is
simplified as
\be \Phi'' + 3 {\cal H} (1 + c_{tot}^2) \Phi' + \Bigl[ 2 {\cal H}'
+ {\cal H}^2(1 + 3c_{tot}^2) \Bigr] \Phi = \frac{a^2}{2\bar{M}^2}
p_{tot} \Gamma_{tot}. \label{metricper3} \ee
If you use the background equations,
this equation can be
 rewritten  as
\be \Phi'' - \Bigl( \ln[\rho_{tot} + p_{tot}] \Bigr)' \Phi' +
\Bigl( \ln[\rho_{tot}/(\rho_{tot} + p_{tot})] \Bigr)' {\cal H}
\Phi = \frac{3 {\cal H}^2}{2} \frac{p_{tot}}{\rho_{tot}}
\Gamma_{tot}. \label{metricper4} \ee If we consider the adiabatic
condition ($p_{tot} \Gamma_{tot} = 0$), then the above equation is
identical to the equation in Bardeen's article~\cite{Bardeen}.As such,
the
 metric perturbation can be rewritten as the curvature
perturbation~\cite{Lyth},
\be \zeta = \frac{({\cal H}^{-1} \Phi' + \Phi)}{(1 +
\omega_{tot})} +  \frac{3}{2} \Phi. \label{zeta} \ee With this we
can express Eq.~(\ref{metricper3}) in the adiabatic
 case as

\be \zeta' = 0. \label{etaprime} \ee
This result looks the same to the
minimally coupled case and there seems to have no difference from the
non-minimally coupled case. Nevertheless, the coupling
information is absorbed in both $c_{tot}^2$ and the perturbation equation
of each species.

\section{Conclusions}
\setcounter{equation}{0}

We have analyzed the linear perturbations of the cosmological
models for the scalar field with its self-interaction potential,
$V(\phi) = V_{0} \exp(\lambda \phi^2 /2)$, and its coupling to the
CDM, $\exp[B_{c}(\phi)] = \exp(n_{c} \lambda \phi^2 /2)$. The
evolution of the non-perturbed background scalar field
$\bar{\phi}$ occurred in the tracking regime throughout the
radiation dominated epoch. The full analysis of the non-perturbed
and perturbed equation of each species in the conformal Newtonian
gauge has been done including the proper energy-momentum transfer
vector due to the coupling, $Q_{(d) \nu}$. The Boltzmann equations
have been modified as to account for the coupling between a scalar
field and
 the CDM.

We have seen that the energy-momentum of each species may not be
conserved as a result of the coupling of the scalar field to the
CDM. Thus we have redefined the concepts of entropy perturbations.
We have shown that the isocurvature perturbation of the
quintessence has been damped to zero with time in the tracking
regime during the radiation epoch. Thus we have constrained our
considerations on the adiabatic perturbation.

We have considered the CMB anisotropy spectrum and the matter
power spectrum for the non-minimally coupled models. Additional
mass and source terms in the Boltzmann equations induced by
the coupling give the rapid changes of the Newtonian potential
$\Phi$ and enhance the ISW effect in the CMB power spectrum. The
modification of the evolution of the CDM, $\rho_{c} =
\rho_{c}^{(0)} a^{-3 + \xi}$, changes the energy density contrast
of the CDM at early epoch. We have adopted the current
cosmological parameters
measured by WMAP within $1 \sigma$ level. With the COBE
normalization and the WMAP data we have found the constraint of
the coupling $n_{c} \leq 0.01$. The locations and the heights of
the CMB anisotropy peaks have been changed due to the coupling.
Especially, there
 is a significant difference for the heights of  the
second and the third peaks among the models. Thus upcoming
observations continuing to focus on resolving the higher peaks may
constrain the strength of the coupling. The suppression of the
amplitudes of the matter power spectra could be lifted by a bias
factor. However, a detailed fitting is beyond the scope of this
paper. The turnover scale of the matter power spectrum may be used
to constrain the strength of the coupling $n_{c}$.

Finally, we have investigated the metric perturbations including
the coupling between the scalar field and the CDM. There is no
difference to the curvature perturbation $\zeta$ for the different
couplings in the adiabatic case. However, the effects of the
coupling have been absorbed in the Boltzmann equations already.

\section{Acknowledgements}
\setcounter{equation}{0}

This work was supported in part by the National Science Council,
Taiwan, ROC under the Grant NSC94-2112-M-001-024 (K.W.N.).


\end{document}